\documentclass[12pt,preprint, epsfig]{aastex}

\shorttitle{Blue stragglers in Hodge 11}
\shortauthors{Chengyuan Li et al.}

\begin{document}

\title{Blue straggler evolution caught in the act in the Large
  Magellanic Cloud globular cluster Hodge 11}

\author{
Chengyuan Li,\altaffilmark{1,2}
Richard de Grijs,\altaffilmark{1}
Licai Deng,\altaffilmark{2} and 
Xiangkun Liu\altaffilmark{1}}

\altaffiltext{1} {Kavli Institute for Astronomy \& Astrophysics and
  Department of Astronomy, Peking University, Yi He Yuan Lu 5, Hai
  Dian District, Beijing 100871, China; joshuali@pku.edu.cn,
  grijs@pku.edu.cn}
\altaffiltext{2} {Key Laboratory for Optical Astronomy, National
  Astronomical Observatories, Chinese Academy of Sciences, 20A Datun
  Road, Chaoyang District, Beijing 100012, China}

\begin{abstract}
High-resolution {\sl Hubble Space Telescope} imaging observations show
that the radial distribution of the field-decontaminated sample of 162
`blue straggler' stars (BSs) in the $11.7^{+0.2}_{-0.1}$ Gyr-old Large
Magellanic Cloud cluster Hodge 11 exhibits a clear bimodality. In
combination with their distinct loci in color--magnitude space, this
offers new evidence in support of theoretical expectations that
suggest different BS formation channels as a function of stellar
density. In the cluster's color--magnitude diagram, the BSs in the
inner 15$''$ (roughly corresponding to the cluster's core radius) are
located more closely to the theoretical sequence resulting from
stellar collisions, while those in the periphery (at radii between
85$''$ and 100$''$) are preferentially found in the region expected to
contain objects formed through binary mass transfer or coalescence. In
addition, the objects' distribution in color--magntiude space provides
us with the rare opportunity in an extragalactic environment to
quantify the evolution of the cluster's collisionally induced BS
population and the likely period that has elapsed since their
formation epoch, which we estimate to have occurred $\sim$4--5 Gyr
ago.
\end{abstract}


\section{Introduction}

In globular cluster (GC) color--magnitude diagrams (CMDs), single
stars that are more massive than the stars located at the
main-sequence turn-off (MSTO) are expected to evolve off the MS. If
all member stars in a GC were to evolve in isolation and without
undergoing any dynamical interactions, the MS locus beyond the MSTO
should be unoccupied. In reality, however, blue straggler stars
(BSs)---which are more massive than stars at the MSTO---nevertheless
occupy the bright MS extension \citep[e.g.,][]{Sand53, Str93}. Two
scenarios for BS formation have been suggested, involving either mass
transfer between or coalescence of close binary companions
\citep[e.g.,][]{McC64, ZS76, Car01, Tian06, Lu2010, Lu2011}, or direct
stellar collisions \citep[e.g.,][]{Hills76, Lomb02, Lu2010, Lu2011}.

The relative importance of these formation channels is still
unclear. \cite{knig09} found a strong correlation between the number
of BSs in Galactic GC (GGC) cores and the associated core mass, but
only a weak correlation with the predicted number of BSs formed
through collisions; such a correlation may only be seen for extremely
dense GGCs, however. Similarly, \cite{Pio04} did not find any
significant correlation between the global BS frequency and the
collision rate in their sample clusters. However, they found that the
BS luminosity function is noticeably different in clusters brighter
and fainter than $M_V=-8.8$ mag \citep[for theoretical arguments,
  see][]{Davies04}. Simulations indicate that the
relative importance of both formation scenarios varies as a cluster
evolves \citep{Hypki12}. 

\cite{Ferr12} suggest that the radial profile of the relative BS
frequency---the number of BSs normalized to that of horizontal-branch
(HB) or red-giant-branch (RGB) stars---may be a good tracer to
quantify a cluster's dynamical age. In dynamically young GCs, where
the relative BS frequency is roughly flat, e.g., $\omega$ Cen, Palomar
14, and NGC 2419 \citep{Ferr12}, BSs formed through mass transfer
dominate \citep[e.g.,][]{Mape06,Hypki12}. On the other hand, in GCs of
intermediate dynamical age the BS frequency's radial profile displays
a `bimodality' \citep{Ferr97,Ferr12,Mape06,Hypki12}. Stellar
collisions dominate the central, crowded region, while mass transfer
dominates the clusters' outer regions (e.g., 47 Tucanae, M3, and NGC
6752; Mapelli et al. 2006). In addition, BSs in dynamically old GCs
(e.g., M30, M75, M79, and M80; Ferraro et al. 2012) exhibit even
clearer signatures of collisional formation. \cite{Ferr09} discovered
two distinct BS sequences in the CMD of M30, which trace the
theoretical single-age stellar-collision and mass-transfer
tracks. \cite{Sills02} showed that collision products are not
chemically homogeneous, so that collisionally formed BSs will be
redder and fainter than their fully chemically mixed counterparts
\citep{Bail95}. This is consistent with the results of \cite{Ferr09}.

One may thus expect for both (dynamically) intermediate-age and old
GCs that the number of BSs formed through either channel should be
comparable. In the absence of any clear substructures, collisionally
induced BSs are expected to be preferentially located in a GC's inner
regions compared with BSs formed through mass transfer. Given the
prediction that BSs of different origins will occupy different parts
of color--magnitude space, the distribution of the different BS types
in the CMD will depend on their radial distance from the cluster
center \citep{Ferr09}. However, since the BSs in a given GC are not
expected to all form simultaneously, the theoretically expected
single-age BS sequences will be broadened \citep[cf.][]{Tian06,
  Lu2010, Lu2011}, and both BS types may partially overlap in
color--magnitude space. However, provided that both populations are
not fully mixed, a robust detection of two subgroups in the CMD should
still be possible.

Studies of extragalactic BSs are relatively rare. \cite{Shara98}
detected 42 BS candidates in the Small Magellanic Cloud cluster NGC
121 and confirmed 23 stars as genuine BSs. In the Large Magellanic
Cloud (LMC) clusters NGC 1466 and NGC 2257, \cite{John99} found 73 and
67 BSs, respectively, while \cite{Mac06}'s analysis of the LMC cluster
ESO 121-SC03 yielded 32 BSs.

Using high-resolution {\sl Hubble Space Telescope (HST)} observations,
here we investigate the radial dependence of the BS frequency in the
LMC GC Hodge 11. We detect a significant number of BSs and a clear
bimodality in the cluster's radial BS frequency. We show that the BSs
in the inner region are located more closely to the theoretical
`collision sequence' (bluer and fainter on average), while those in
the cluster's outskirts are preferentially found in the `binary
region' (redder and brighter). This result, combined with the
discovery that Hodge 11 hosts one of the largest BS populations in any
Galactic or extragalactic cluster known to date,\footnote{Although
  $\omega$ Cen and NGC 2419 host 313 and 266 BSs, respectively
  \citep{ferr06,dales08}, whether or not these objects are genuine GCs
  is as yet unresolved \citep[e.g.,][and references
    therein]{mackey05}.} offers unprecedented evidence in support of
the theoretical predictions and allows us to quantify the evolution of
the cluster's collisionally formed BS population.

\section{Data}

\begin{figure}[ht!]
\plotone{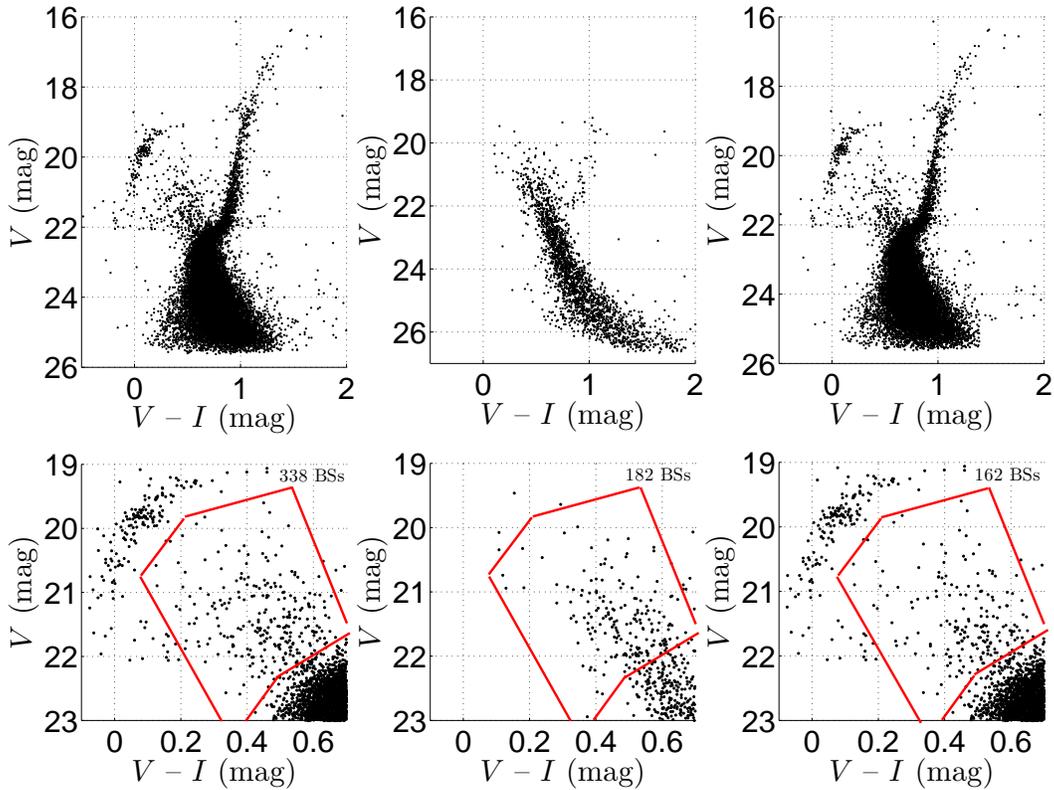}
\caption{(top left) Original Hodge 11 CMD. (top middle) CMD of a
  carefully selected nearby field region \citep[cf.][]{Castro01}. (top
  right) Hodge 11 CMD after field-star decontamination. (bottom)
  Corresponding CMD zooms focusing on the BS-dominated region. The
  number of BSs and the boundaries adopted for the BS region (red
  outlines; see text) are also indicated. The small difference between
  the total number of objects in the BS region (bottom left) and the
  sum of the BSs identified in the bottom-middle and bottom-right
  panels is due to the statistical nature of our field-star
  decontamination procedure \cite[cf.][]{Hu10}.}
\label{FigA}
\end{figure}

\begin{figure}[ht!]
\plotone{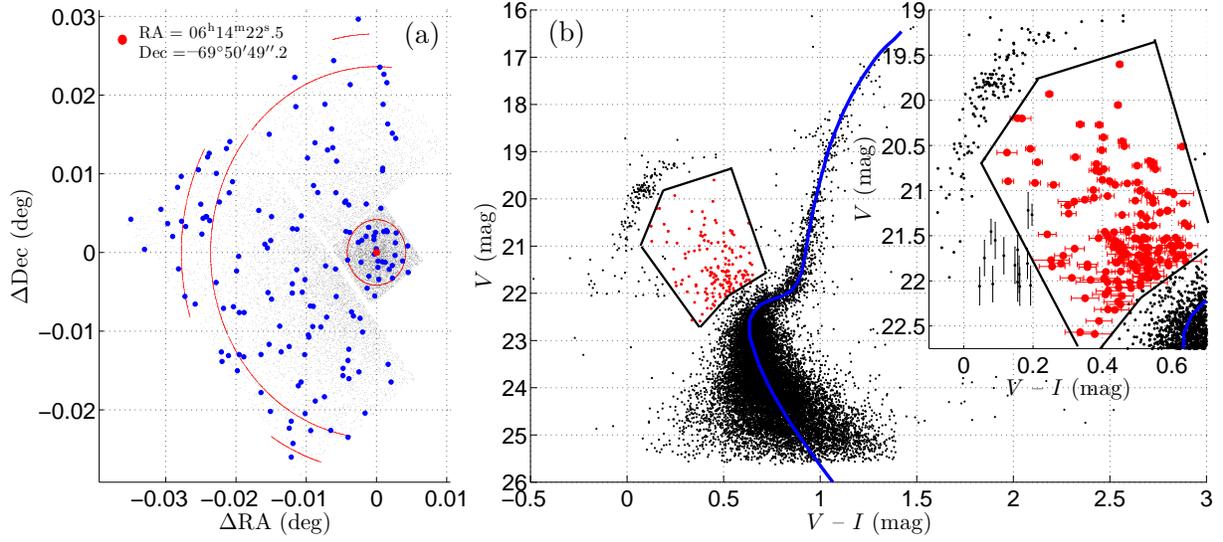}
\caption{(left) Spatial distribution of stars observed in the Hodge 11
  area. Red point: Cluster center. Blue points: BSs. Red (partial)
  concentric circles indicate radii of $R = 15, 85$, and $100''$.
  (right) Corresponding {\sl HST} CMD, including the best-fitting
  \cite{Bert08} isochrone (blue). Red points represent our BS
  selection; the inset shows the photometric uncertainties for all BS
  candidates. Black points with error bars have photometric errors
  $\Delta V > 0.1$ mag (color uncertainties have been omitted for
  clarity). Thick solid lines delineate the BS region adopted.}
\label{fig1}
\end{figure}

Our data sets were obtained with the {\sl HST}'s Wide Field and
Planetary Camera-2, employing the F555W (equivalent and henceforth
referred to as $V$) and F814W ($\sim I$) filters \citep[for
  observational details, see][]{grijs02}. The total exposure times of
the observations used were 435 and 960 s in the $V$ and $I$ filters,
respectively. We applied point-spread-function fitting to obtain the
cluster's stellar photometric catalog \citep{Dol2000a,Dol2005}. The
cluster center was determined as described by \cite{degrijs13}. We
also obtained the stellar catalog of a nearby field region \citep[for
  observational details, see][]{Castro01} to statistically reduce
contamination by background stars \citep[for procedural details,
  see][]{Hu10}: the results are shown in Fig. \ref{FigA}. The number
of objects located in the BS region (see below for definition) in the
original Hodge 11 CMD is 338; the decontaminated CMD contains 162
stars in the same region. This latter sample forms the basis of our
analysis in this {\it Letter}. Our photometric catalog is $> 95$\%
complete in the relevant magnitude range, even in the highest-density
central region. Figure \ref{fig1}a displays the spatial distribution
of all observed stars in the Hodge 11 area covered by our {\sl HST}
field of view, as well as its BS population.

We divided the MS and RGB regions in the Hodge 11 CMD into 32 bins of
$\sim 0.30$ mag each between $V=18.3$ and 27.0 mag. In each magnitude
range, we determined the peak and $3 \sigma$ spread of the color
distribution using a Gaussian function (which provided appropriate
fits at any magnitude). Spectroscopic abundance determinations show
that Hodge 11 is metal-poor, [Fe/H] = $-2.1\pm0.2$ dex
\citep{Cow82,Ols91} to $-1.86$ dex \citep{Gro06}. Adopting a
metallicity $Z=0.0002$ ([Fe/H] $\approx-1.98$ dex) and using the
ridgeline thus obtained, we determined the best-fitting isochrone
based on the Padova stellar evolutionary models \citep{Bre12} and
calculated the MSTO locus. We thus determined a cluster age of $\log(t
\mbox{ yr}^{-1})=10.07 \pm 0.01$ \citep[$11.7^{+0.2}_{-0.1}$ Gyr;
  cf.][]{John99} and an extinction of $E(B-V)=0.09\pm0.01$ mag. The
latter is in excellent agreement with \cite{Wal93}'s determination of
$E(B-V)=0.08\pm0.02$ mag. We adopted the canonical LMC distance
modulus, $(m-M)_0=18.50$ mag, so that $1'' \equiv 0.24$ pc.

Many of the cluster's HB stars are located in the extreme blue HB
(EHB). Our sample BSs can easily be distinguished from these (E)HB
stars by adopting a minimum (E)HB--BS difference in color and
magnitude of 0.3 mag. Our selection is robust, since the (E)HB stars
and BSs are clearly separated in the CMD (see Fig. \ref{fig1}) and the
absolute photometric errors are $\ll 0.3$ mag.

Finally, we identify a star as a likely BS if it meets all of the
following criteria:
\begin{enumerate}
\item It is bluer and brighter than the single-star MSTO locus.
\item Its distance in color--magnitude space from the MSTO and MS
  ridgeline is $> 3\sigma$ ($\sigma$ refers to the intrinsic spread in
  MS color at the magnitude of interest).
\item Its photometric uncertainty is $< 0.1$ mag in both filters.
\end{enumerate}

Figure \ref{fig1}b shows the Hodge 11 {\sl HST} CMD, where the inset
focuses on the region typically populated by BSs. The red points
represent our (field-decontaminated) BS candidates. The small black
points with error bars are objects whose photometric errors are $>
0.1$ mag, while most of the error bars of the red points are smaller
than the symbol size. Inspection on the original images of the sources
with large error bars revealed that they either are affected by
instrumental artefacts (including low signal-to-noise ratios or
closely juxtaposed to bad pixels) or may be unresolved blends. We thus
discarded these objects from our BS sample.

\section{Analysis and results}

\begin{figure}[ht!]
\plotone{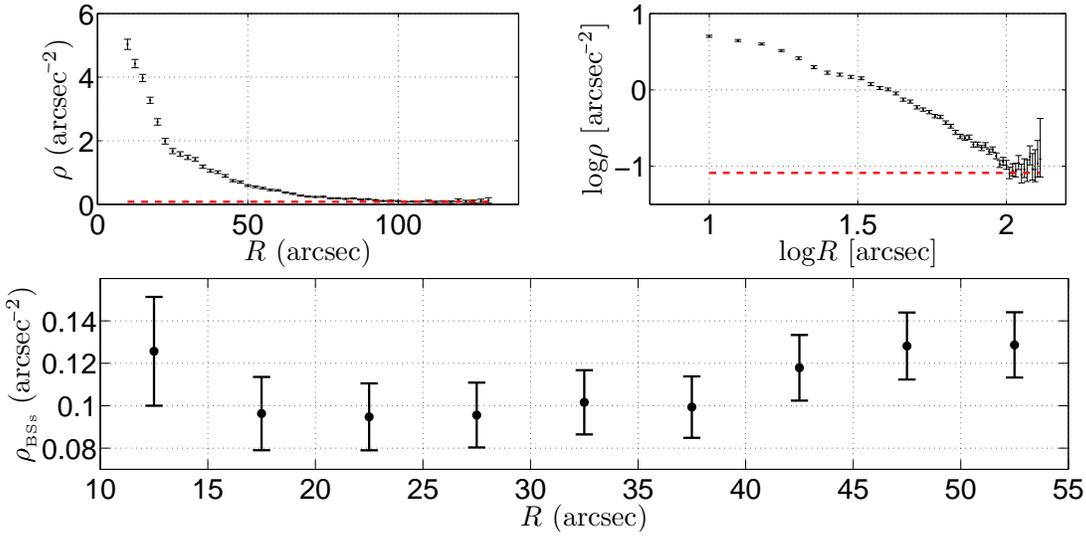}
\caption{(a) Hodge 11 number-density profile (corrected for partial
  areal coverage). The red dashed line indicates the average field
  density, implying a cluster size of $R = 102.3^{+2.7}_{-2.3}$
  arcsec. (b) Corresponding logarithmic number-density profile. (c)
  BS-fraction profile (normalized to the number of RGB stars) of Hodge
  11, including Poissonian uncertainties.}
\label{fig2}
\end{figure}

Figures \ref{fig2}a and \ref{fig2}b show the cluster's stellar number
density as a function of radius for all stars with $V \le 25$ mag.
Based on the catalog of the nearby field region, we calculated the
field's average density value. We adopted the distance where the
cluster's monotonically decreasing number-density profile reaches the
average field level, $R=102.3^{+2.7}_{-2.3}$ arcsec, as the cluster
region.

In addition to our BS sample, we selected a sample containing RGB
stars covering the magnitude range from $V = 19.60$ to 22.05 mag. We
normalized the number of BSs to the number of RGB stars: $f_{\rm BSs}
= N_{\rm BSs}/N_{\rm RGB}$, where $N_{\rm BSs}$ and $N_{\rm RGB}$
represent the number of BSs and RGB stars, respectively. Previous
studies \citep[e.g.,][]{Ferr03} normalized the number of BSs to either
or both HB or/and RGB stars. We adopted RGB stars as reference for
reasons of consistency with previous analyses. We subsequently
determined the radial dependence of the BS fraction, $f_{\rm BSs}(R)$:
see Fig. \ref{fig2}c. Normalization of the BS number to the number of
HB stars results in a similar, although somewhat flatter trend. This
BS-fraction profile is very similar to the equivalent radial
dependences in 47 Tuc, M3, and NGC 6752 \citep{Mape06}.

The number of BSs in Hodge 11 is extremely large. \cite{Ferr95}
summarized the number of BSs detected in 26 GGCs, ranging from four
(NGC 5024) to 137 (NGC 5272). In Hodge 11, we detected 150 BSs within
$R = 100''$;  
the total number of BSs on all four chips of our Hodge 11 observations
reaches 162 (see Section \ref{disc.sec} for an assessment of the level
of completeness of our observations' areal coverage). Since the BS
distribution associated with Hodge 11 is very extended, the number of
BSs located in our outer subsample, $R\in[85'',100'']$, is
significant. This annulus contains 27 BSs, and hence this large number
of BSs strengthens the statistical reliability of our conclusions.

\begin{figure}[ht!]
\plotone{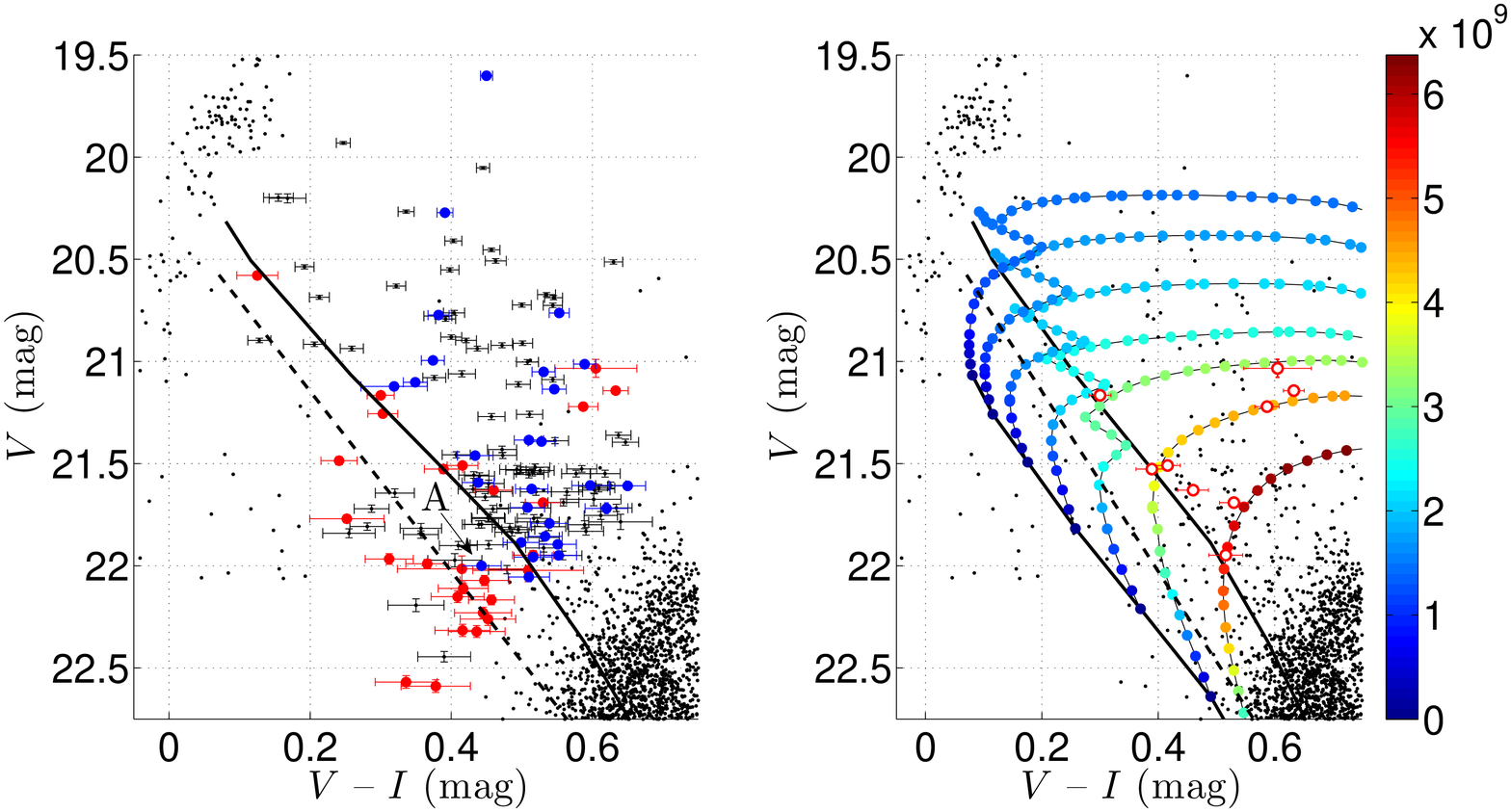}
\caption{(left) Color--magnitude distribution of BSs in Hodge 11. Red
  points: BSs at $R\le 15''$; blue points: BSs at
  $R\in[85'',100'']$. The black dashed line indicates approximately
  the critical line separating the two BS sequences in M30
  \citep{Ferr09}; the black solid line indicates the predicted ZAMS
  adjusted by $-0.75$ mag. (right) Evolutionary tracks of stars with
  masses from 1.0 to $1.6M_{\odot}$ in steps of $0.1 M_{\odot}$
  (bottom to top; Bertelli et al. 2008) for $Z = 0.0004$ (closest
  available metallicity). Different colors indicate the evolutionary
  timescales for stars to evolve from the ZAMS to their current
  positions, in units of $10^9$ yr. Red open circles: Inner-sample BSs
  ($R\le 15''$) located above the B-ZAMS.}
\label{fig3}
\end{figure}

\cite{Mac03} determined a cluster core radius of $R_{\rm core} =
12.14\pm0.66$ arcsec. Our number-density profile exhibits a relatively
sharp radial decline to $R \sim 15$--$20''$. Therefore, we defined an
inner BS subsample covering the cluster core at $R\le 15''$
(Fig. \ref{fig3}, left). The red and blue points in the CMD of
Fig. \ref{fig3} (left) represent the inner ($R\le 15''$) and outer
($R\in[85'',100'']$) subsamples. Both contain 27 BSs. A distinct
difference is seen between the mean loci of both samples in
color--magnitude space. We also show the separation between the two BS
subpopulations in M30 found by \citep[][dashed line]{Ferr09}. These
authors suggested that BSs below this `critical' line might
preferentially result from stellar collisions in relatively
high-density regions, while BSs above the line are predominantly
thought to have formed through mass transfer between binary companions
in lower-density areas.

Considering that M30 and Hodge 11 may have different evolutionary
histories and environments, using the M30 critical line to analyze the
color--magnitude distribution of the Hodge 11 BSs may not be
appropriate. If we shift the zero-age main sequence (ZAMS) upward in
the CMD by 0.75 mag---which represents the ZAMS of equal-mass binary
systems (B-ZAMS)---we see that this provides an excellent separation
between both subsamples. Hence, our results for Hodge 11 support the
claims of \cite{Ferr09} for M30: Only one outer-sample BS is
unambiguously located in the `bottom region' (labeled `A' in
Fig. \ref{fig3}, middle), while the 26 remaining outer-sample BSs are
all located above the B-ZAMS (modulo the photometric
uncertainties). In contrast, for our $R \le 15''$ inner-sample BS
selection, only four BSs are unambiguously located in this CMD
region, given the photometric errors.

\section{Discussion}
\label{disc.sec}

Since the cluster's central region is very crowded, the BSs in this
region are most likely preferentially the result of stellar
collisions. On the other hand, since the low number density at large
radii renders a high frequency of direct stellar collisions unlikely,
the BSs in the cluster's periphery most likely originate from mass
transfer between the individual members in binary systems. Based on
the B-ZAMS adopted, the mixture of the two BSs subsamples in the CMD
seems `unbalanced:' only few outer-sample BSs are found in the bottom
region, while inner-sample BSs frequently `pollute' the top region. We
suggest three possible explanations for this observation:
\begin{enumerate}
\item The spatial distribution of the BSs in Hodge 11 is a
  two-dimensional projection of a 3D spatial distribution. Some
  apparent inner-sample BSs may not be genuine inner-sample members,
  but could be outer-sample BSs that are projected close to the
  cluster center. On the other hand, outer-sample BSs must always be
  located truly far away from the center. Geometry considerations,
  assuming a uniform BS distribution in the outer annulus, $R\in[85'',
    100'']$, suggest that $4 \pm 2$ (Poissonian uncertainty)
  outer-sample BS may be projected along the full line of sight
  through the cluster onto radii $R \le 15''$.

\item BSs evolve and trace paths in the CMD. Without any new dynamical
  interactions, the BSs in the bottom region will move upward, but the
  BSs in the top area will never move downward. To estimate the
  approximate time necessary for BSs to evolve from the ZAMS to the
  region above the B-ZAMS, in Fig. \ref{fig3} (right) we show
  evolutionary tracks from \cite{Bert08} for stars of different
  masses. Below, we select those inner-sample BSs that are located
  above the B-ZAMS (red open circles) and compare their loci to the
  tracks for stars with masses from 1.0 to $1.6M_{\odot}$ (bottom to
  top; $1.6M_{\odot}$ is approximately $2 M_{\rm MSTO}$). Different
  colors indicate different evolutionary timescales for stars to
  evolve from the ZAMS to their current positions.
  
\item There may still be some primordial binaries in the inner region,
  which may merge to form BSs. Once formed, they will be located in
  the top region.
\end{enumerate}

One outer-sample BS, object `A' in Fig. \ref{fig3} (middle), is found
in the CMD's bottom region. Its locus cannot have been caused by the
expected evolutionary color spread. This star could either be a
projected foreground star or a collisionally formed cluster BS that
may have been ejected dynamically to the cluster's periphery
\citep[cf.][]{Sig94}. Such ejected BSs will subsequently sink to the
center again through the effects of two-body relaxation (dynamical
mass segregation). Compared with the other cluster BSs, this star is
relatively faint (and hence of relatively low mass), so that this
relaxation process is likely still underway.

Accounting for scenarios 1--3 above, the radial distance bias of BSs
in the Hodge 11 CMD is nevertheless significant, which suggests that
our result is robust. In addition, compared with M30, the number of
{\it field} BSs detected in the Hodge 11 field at radii beyond the
cluster's outer radius is very large. We identified 12 BSs in the
partially observed annulus $R\in[100'', 130'']$, whose {\sl HST}-based
coverage is only 9.3\% complete (for comparison, our coverage of the
actual cluster area, $R \le 100''$, is 50.2\% complete). Nevertheless,
if we adopt an outer boundary to the outer-sample BS population such
that all observed BSs are included and compare their color--magnitude
distribution with that of the inner-sample BSs, the distinction
between the two samples remains significant.

If we treat all selected objects as genuine cluster BSs, the locus of
the most massive BS---which should originally have been located in the
`bottom region' of the CMD but is found above the B-ZAMS---may have
been caused by stellar evolution. Geometry considerations imply that
$4 \pm 2$ stars may be seen in projection, although nine inner-sample
BSs that may have crossed the B-ZAMS are actually detected. This means
that $5 \pm 2$ objects could be evolved BSs. If correct, this allows
us to estimate their time of birth at 4--5 Gyr ago. The BSs
characterized by magnitudes between $V=21.0$ and 21.5 mag are probably
also evolved stars. For BSs fainter than $V=21.5$ mag, the time
required to evolve to their current loci is so long that their
positions in the CMD may be owing to 2D projection of the cluster's 3D
spatial distribution.
    
\section{Conclusion}

Based on the predictions of stellar evolutionary models, BSs resulting
from direct stellar collisions and from mass transfer between binary
companions will be located along two distinct sequences. However, two
well-separated sequences can only be observed if all BS formation in a
given cluster occurred recently or if all BSs are
short-lived. Long-term, continuous BS formation will render the
expected boundary less distinct. Nevertheless, it is expected that BSs
formed through either of these two mechanisms show the signature of a
clustercentric radial dependence in their color--magnitude
distribution. We analyzed the {\sl HST} CMD of the GC Hodge 11, where
we found a clear signature of such a spatial bias. The average BS in
the inner region is relatively faint, while BSs in the outer region
are brighter. Even when considering the possible effects of projection
and evolution, this difference is robust. Our result hence offers
strong evidence in support of the theoretical expectation of dual-mode
BS formation and allowed us to quantify their evolutionary timescales.

\section*{Acknowledgements}

We thank Peter Anders and the referee for useful suggestions. We are
grateful for support from the National Natural Science Foundation of
China through grants 11073001 and 10973015.

\end{document}